\documentclass[conference, letterpaper]{IEEEtran}
\IEEEoverridecommandlockouts
\usepackage{cite}
\usepackage{amsmath,amssymb,amsfonts}
\usepackage{algpseudocode}
\usepackage{algorithm}
\usepackage{graphicx}
\usepackage{textcomp}
\usepackage[table]{xcolor}
\usepackage{multirow}
\usepackage{url}
\usepackage{listings}
\usepackage{makecell}
\usepackage{color}
\usepackage{hyperref}
\usepackage[english]{babel}
\def\BibTeX{{\rm B\kern-.05em{\sc i\kern-.025em b}\kern-.08em
    T\kern-.1667em\lower.7ex\hbox{E}\kern-.125emX}}
\definecolor{codegreen}{rgb}{0,0.6,0}
\definecolor{codegray}{rgb}{0.5,0.5,0.5}
\definecolor{codepurple}{rgb}{0.58,0,0.82}
\definecolor{backcolour}{rgb}{0.95,0.95,0.92}
\lstdefinestyle{mystyle}{
    backgroundcolor=\color{backcolour},
    commentstyle=\color{codegreen},
    keywordstyle=\color{magenta},
    numberstyle=\tiny\color{codegray},
    stringstyle=\color{codepurple},
    basicstyle=\ttfamily\footnotesize,
    breakatwhitespace=false,
    breaklines=true,
    captionpos=b,
    keepspaces=true,
    numbers=left,
    numbersep=5pt,
    showspaces=false,
    showstringspaces=false,
    showtabs=false,
    tabsize=2
}
\begin{document}
\title{Quantum Software Engineering in Practice: FPGA and AI Integration for Quantum Certification}

\author{
\IEEEauthorblockN{Marcos Guillermo Lammers\IEEEauthorrefmark{1} \quad José Manuel Suárez\IEEEauthorrefmark{1}}
\IEEEauthorblockA{\IEEEauthorrefmark{1}\textit{PhD Candidate in Computer Science},
\textit{Facultad de Informática, LIFIA},
\\
\textit{Universidad Nacional de La Plata}, La Plata, Argentina \\
marcos.lammers@lifia.info.unlp.edu.ar \quad jsuarez@lifia.info.unlp.edu.ar}
\\
\IEEEauthorblockN{Adrián Pousa\IEEEauthorrefmark{2}}
\IEEEauthorblockA{\IEEEauthorrefmark{2}\textit{Facultad de Informática, III-LIDI},
\textit{Universidad Nacional de La Plata}, La Plata, Argentina \\
apousa@lidi.info.unlp.edu.ar}
\\
\IEEEauthorblockN{Luis Mariano Bibbó\IEEEauthorrefmark{3} \quad Alejandro Fernández\IEEEauthorrefmark{3}}
\IEEEauthorblockA{\IEEEauthorrefmark{3}\textit{Facultad de Informática, LIFIA},
\textit{Universidad Nacional de La Plata}, La Plata, Argentina \\
lmbibbo@lifia.info.unlp.edu.ar, alejandro.fernandez@lifia.info.unlp.edu.ar}
}

\maketitle

\begin{abstract}

The emergence of Quantum Software Engineering (QSE) responds to the need for systematic, disciplined, and quantifiable approaches to the development, operation, and maintenance of quantum software. Within this context, quantum computer certification represents a challenge: how to verify that quantum devices produce valid entangled states despite hardware imperfections, noise, and decoherence. This paper presents QAccCert, a hybrid certification framework developed following QSE principles, which demonstrates how heterogeneous technologies, specifically FPGAs and Artificial Intelligence, can be integrated for quantum processing. The framework implements entanglement certification through CHSH inequality violation in ideal quantum simulations (Qiskit AerSimulator), achieving 99.94\% of the theoretical limit ($2\sqrt{2}$) through LLM guided optimization, evidencing more efficient parameter space exploration than random search. These simulated results illustrate how QSE methodologies, combined with strategic technology interconnection, can be used for practical and scalable quantum certification when applied to real NISQ hardware in future work. This work provides a concrete case study of systematic quantum software development.
\end{abstract}


\section{Introduction}
Quantum computing has advanced significantly over the last decade, giving rise to the NISQ (Noisy Intermediate-Scale Quantum) era, characterized by devices with hundreds of qubits but still prone to errors and subject to serious operational limitations. This scenario has driven the emergence of a new discipline: Quantum Software Engineering (QSE), which seeks to apply systematic, disciplined, and quantifiable approaches to the development, operation, and maintenance of quantum software and the hybrid quantum-classical systems with which it integrates.
Among the challenges currently addressed by QSE is the systematic and automated certification of quantum computers—that is, verifying that they operate correctly and produce valid quantum states. Building on previous work \cite{Holik2024GroupinvariantEO}, it is evident that this is a critical problem involving aspects such as:
\begin{itemize}
    \item NISQ devices are inherently noisy and error-prone.
    \item The degree of entanglement of quantum states, which represents a vital resource in the characterization of quantum computers.
    \item Hardware imperfections, such as decoherence, asymmetric noise, calibration errors, and thermal drift, invalidate the ideal assumptions of theoretical protocols.
\end{itemize}
Traditionally, quantum certification and benchmarking protocols have been addressed through methods that assume ideal conditions or do not broadly address NISQ imperfections from a software engineering perspective \cite{Quetschlich2023mqtbench, benchq_github, rigetti_github}. However, work from the field of quantum physics \cite{Arango, Holik2024GroupinvariantEO} and QSE practice teaches us that software systems interacting with imperfect hardware require hybrid approaches that coherently integrate multiple technologies.

\subsection{Technology Interconnection as a Response}
This paper argues that the response to many of the challenges posed by quantum hardware certification must necessarily occur within a hybrid quantum-classical context, placing the focus on the strategic interoperability of emerging technologies guided by QSE principles \cite{ashis2025, Murillo2025}, such as the following:
\begin{itemize}
    \item \textbf{Dedicated hardware (FPGAs)}, to accelerate the processing of quantum correlations. In particular, we use an open-hardware FPGA, such as the Kéfir project \cite{fpgalibre_kefir} and the visual environment Icestudio \cite{fpgaWars_icestudio}, as they allow broader access to this technology.
    \item \textbf{Artificial Intelligence (LLMs)}, to adaptively optimize measurement parameters, selected for their open access and balance between capability and computational efficiency.
    \item \textbf{Standardized quantum protocols}, for state generation and measurement \cite{Arango, Holik2024GroupinvariantEO}.
    \item \textbf{Scalable hybrid infrastructure}, serving as a constructive foundation for managing the enormous volume of data in post-processing stages, as well as the appropriate management of classical quantum resources \cite{SHEHATA2026, BECK202411, Saurabh2023}.
\end{itemize}
Framed within a properly defined software architecture and infrastructure, this facilitates a certification process that is scalable, hardware-agnostic, and aligned with the foundations of QSE \cite{aparicio-morales2024}.

\subsection{Related Work and Context}
In recent years, various proposals have emerged addressing the limitations of NISQ devices from different perspectives, evidencing the growing need for tools that automate optimization and hardware adaptation. In the academic domain, analyses have been developed on quantum resource management \cite{lammers2025gestion} and its extension in an experimental framework called \textbf{Qonscious} \cite{lammers2025qonscious}, which proposes the conditional execution of programs based on the dynamic availability of resources. In parallel, commercial solutions such as \textbf{Fire Opal} by Q-CTRL \footnote{Q-CTRL - \url{https://q-ctrl.com/}} \cite{qctrl_fireopal} offer a hardware abstraction layer that integrates automatic error suppression and circuit optimization, allowing users to run algorithms on multiple backends without reconfiguration and improving the scalability of results.
In the area of resource estimation and efficient simulation, frameworks such as those offered by Zapata Computing (BenchQ) \cite{benchq_github} and Rigetti (Rigetti Resource Estimation) \cite{rigetti_github} use advanced techniques such as stabilizer circuit simulation based on graph state representations \cite{PhysRevA.73.022334} and circuit decomposition methods for fault-tolerant resource estimation \cite{2209.07345}. However, these tools do not explicitly account for the imperfections and dynamic adaptation required in the NISQ era.
In this work, we highlight the need to combine specialized hardware, artificial intelligence, infrastructure, and high-level software to overcome the limitations of NISQ devices. Our accelerated quantum certification prototype (\textbf{QAccCert}) distinguishes itself by specifically addressing the problem of quantum entanglement certification through the synergistic combination of FPGA acceleration and AI optimization via LLMs. Deploying an integrated hybrid prototype solution would be applicable to a broad spectrum of quantum infrastructures and technologies, flexibly meeting the dynamic requirements of each algorithm through software abstraction layers that enable this decoupling.
Considering the context and objective of this work, it is necessary to contemplate a supporting software architecture that enables the integration of the potential of quantum computing with HPC infrastructures in pre- and post-quantum processing stages. Such a hybrid architecture must be hardware-infrastructure-agnostic and adaptable to both current NISQ computers and future FTQC (Fault-Tolerant Quantum Computing) systems. In this regard, a framework is proposed for the efficient management of data between quantum and classical resources, as well as a proposed technology stack in \cite{SHEHATA2026}, where the key element is a QGateway that efficiently manages resource allocation.
An analysis is also conducted on application requirements based on the variable level of quantum resource demand. Tasks such as quantum benchmarking, the central use case of QAccCert, would require low QC load but many requests, in what might be seen as frequent and fragmented interaction, and also high HPC demand in relation to interaction with LLMs and results management in post-processing stages.
Regarding the challenges posed by QC-HPC integration, we consider resource management, job scheduling, efficient data flow management, hardware and infrastructure independence, service discovery, interconnection networks, and exposure of quantum parameters.
The work of \cite{BECK202411} addresses system integration through quantum computational acceleration on HPC environments via a hardware-agnostic framework. This integration is developed from the perspective of hardware, users, applications, software, and development flow. Notably, the authors subdivide the component integration scheme into `tightly coupled' or `loosely coupled', depending on whether the CPU-GPU-QPU triad can or cannot be managed under the same workflow.
Finally, the work of \cite{Saurabh2023} proposes a conceptual middleware that identifies the particularities of the quantum-classical integration process, sub-classifying the different possible interaction scenarios into: HPC-for-quantum, quantum-in-HPC, and quantum-on-HPC, primarily considering the level of coupling and the structure of the application, as well as their impact on middleware requirements. These works demonstrate that QC-HPC integration is an active and multidimensional challenge, whose systematic treatment constitutes one of the axes on which QAccCert builds its architecture.

\section{Contributions and Structure}
To substantiate our hypothesis, we developed QAccCert (Quantum Accelerated Certification Framework), with a hybrid approach following QSE guidelines that implements quantum entanglement certification through violation of the CHSH inequality \cite{Arango}. QAccCert serves as a case study and aims to demonstrate:
\begin{enumerate}
    \item How QSE can guide the integration of heterogeneous technologies.
    \item How FPGA acceleration significantly reduces classical post-processing times.
    \item How LLM-based optimization enables more efficient exploration of the parameter space within specific hardware imperfections.
    \item How a hybrid architecture incorporating HPC can handle data volumes and post-processing requirements.
    \item The feasibility of these approaches in real-world scenarios.
\end{enumerate}
The paper is structured as follows: Section II analyzes the problem of quantum certification from the perspective of Quantum Software Engineering, emphasizing why traditional methods are often insufficient and why the FPGA+AI combination is advantageous, as well as a prototype of a supporting architecture. Section III presents the proposed technology interconnection architecture, with QAccCert as a concrete implementation. Section IV shows the experimental results obtained, validating the approach. Finally, Section V discusses the implications for QSE, explores possible future research directions, and summarizes the conclusions.

\section{Quantum Certification as a Quantum Software Engineering Problem}
\subsection{What Makes Certification a QSE Challenge?}
Quantum Software Engineering is concerned with ``systematic, disciplined, and quantifiable approaches to the development, operation, and maintenance of quantum software.'' Quantum computer certification encompasses all of these characteristics:
\begin{itemize}
    \item \textbf{Systematic}: It requires well-defined protocols (such as CHSH) that must be executed in a reproducible manner.
    \item \textbf{Disciplined}: It demands the establishment of adequate metrics, as well as rigorous quality control and validation processes.
    \item \textbf{Quantifiable}: It produces numerical values (violation $S$) that allow for the objective evaluation of the system's state.
    \item \textbf{Operation and maintenance}: Certification is not a one-time event, but a continuous process that must adapt to changes in hardware (thermal drift, aging, recalibration, etc.).
\end{itemize}
Furthermore, certification involves hybrid software that must:
\begin{itemize}
    \item \textbf{Interact} with quantum hardware or software when working with simulation, typically through APIs such as Qiskit \footnote{IBM-Qiskit - \url{https://www.ibm.com/quantum/qiskit}}.
    \item \textbf{Process and manage} the large volumes of measurement data generated.
    \item \textbf{Be flexible and adaptive} to changing conditions in real time.
    \item \textbf{Integrate and interoperate} with classical control and acquisition systems.
\end{itemize}

\subsection{Limitations of Traditional Approaches from the QSE Perspective}
Certification based on Bell inequality violation typically uses the CHSH inequality \cite{Arango}:

\begin{equation}
S = |E(a,b) - E(a,b') + E(a',b) + E(a',b')| \leq 2
\label{eq:chsh}
\end{equation}
where $E(\alpha,\beta)$ represents the correlation between measurements in bases $\alpha$ and $\beta$. The theoretical quantum limit is $2\sqrt{2} \approx 2.828$.
The theoretically optimal measurement angles ($\theta = [a,a',b,b']$) have a known analytical solution. However, from a QSE perspective, multiple problems arise:
\begin{itemize}
    \item \textbf{Hardware-software coupling}: The theoretically optimal angles assume ideal hardware; in practice, each device has unique imperfections that require software adaptation.
    \item \textbf{Data quality}: Noise and decoherence introduce uncertainty into measurements.
    \item \textbf{Computational efficiency}: The search for optimal experimental angles is costly if performed by brute force.
    \item \textbf{Maintainability}: A certification framework must be updatable when hardware changes or new protocols emerge.
    \item \textbf{Reproducibility}: Results must be consistent across runs and, ideally, across different installations.
\end{itemize}
As a result, the theoretically optimal angles do not coincide with the experimentally optimal angles, and the actual violation is lower than expected. Validated certification requires finding these experimental optima. This is a task that demands a software engineering approach to be practical.

\subsection{Why FPGA and AI? A Response from QSE}
QSE teaches us that complex systems require modular architectures where each component specializes in a function. In our case:
\begin{itemize}
    \item \textbf{FPGA as a hardware accelerator}: FPGAs offer parallel processing of multiple quantum correlations, optimized integer arithmetic, and low latency. From the QSE perspective, this is a design pattern for separating intensive computation from high-level control. This approach is complementary to others such as the automatic error suppression proposed by commercial solutions \cite{qctrl_fireopal} or resource-based conditional execution \cite{lammers2025gestion}.
    \item \textbf{AI (LLMs) as an adaptive optimizer}: LLMs can identify patterns in historical data and suggest promising configurations. In QSE terms, this constitutes a learning and adaptation component that improves with experience, similar in spirit to the resource management analyses proposed in \cite{lammers2025gestion} and the symmetry-invariant estimation methods developed in \cite{Holik2024GroupinvariantEO}.
    \item \textbf{Standardized quantum protocols}: State preparation and measurement follow defined interfaces (Qiskit), enabling interchangeability of the quantum backend.
\end{itemize}
The central thesis of this work is that the synergistic combination of these technologies, framed within a well-designed QSE architecture, constitutes in itself the main contribution: the numerical results presented later serve as evidence of its coherence and feasibility. This integration scales as algorithms and topologies grow, being applicable to both pre- and post-processing of data.

\section{Technology Interconnection Architecture: QAccCert as a QSE Case Study}
\begin{figure}[htbp]
\centerline{\includegraphics[width=1\columnwidth]{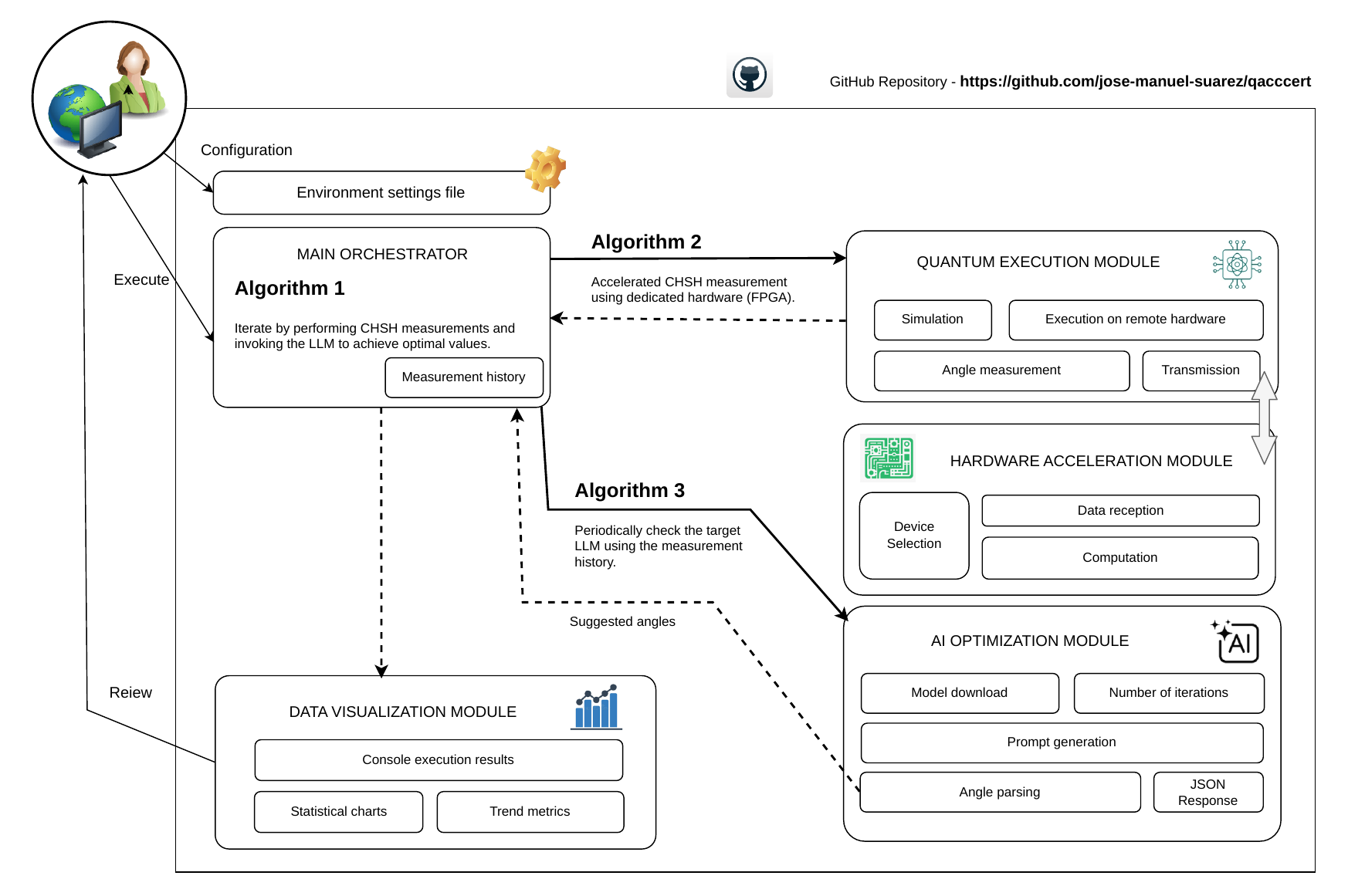}}
\caption{Hybrid architecture of QAccCert.}
\label{fig:arquitectura}
\end{figure}

\textbf{QAccCert} (an acronym for \textbf{Q}uantum \textbf{Acc}elerated \textbf{Cert}ification Framework) implements a hybrid optimization cycle following Quantum Software Engineering principles: separation of concerns, defined interfaces, and reusable components. An illustrative diagram of the proposed architecture is shown in Figure~\ref{fig:arquitectura}.

\subsection{Architecture Components}
\begin{enumerate}
    \item \textbf{Quantum state preparation module}: Parameterized circuits generate Bell states, see Figure~\ref{fig:circuito}. This module abstracts the underlying hardware (simulator or real device) through a unified interface (Qiskit).
  
    \begin{figure}[htbp]
    \centerline{\includegraphics[width=0.6\columnwidth]{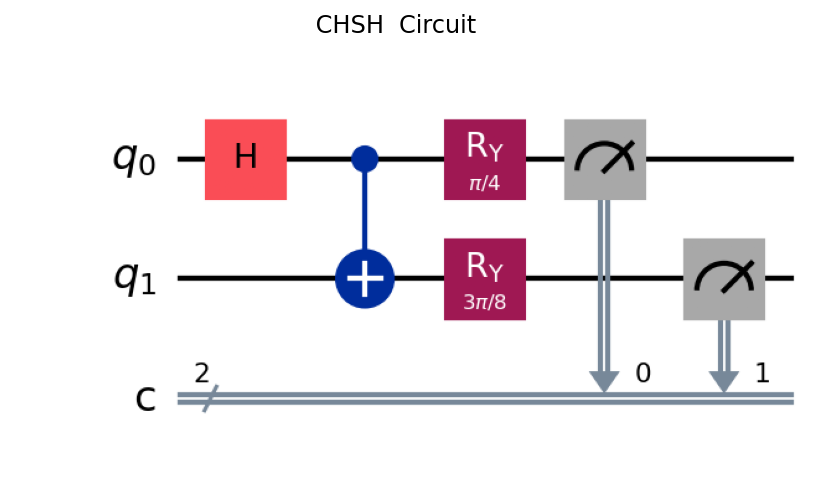}}
    \caption{Quantum circuit for measuring non-locality in entangled states with some of the Ry gate rotations.}
    \label{fig:circuito}
    \end{figure}
  
    \item \textbf{FPGA-accelerated processing module}: Raw measurements were transmitted using the UART protocol so that the FPGA computes the correlations $E(\alpha,\beta)$. In our prototype, it was programmed in Verilog code and the FPGA synthesis was performed using the visual environment Icestudio \cite{fpgaWars_icestudio}, an open-source tool that simplifies digital design for open-hardware FPGAs. The FPGA board was developed based on a design from the Kéfir project \cite{fpgalibre_kefir}. The specific integrated circuit (FPGA) is manufactured by Lattice Semiconductor, iCE40 Ultra family, model iCE40HX4K-TQ144, with 40 nm technology, 3520 logic cells (4-input LUTs), 80 Kib of distributed SRAM, 2 PLLs for clock management, 107 I/O pins at 3.3 V; 4 Mib SPI Flash configuration memory (Winbond W25X40CL), with approximately 3 Mib available for data storage. This module encapsulates the low-level logic and exposes a simple API to the rest of the system.
  
    \item \textbf{AI-based optimization module}: For the analysis of the iteration history and suggestion of promising angles, we selected two representative large language models (LLMs): Mistral-7B-Instruct-v0.2 \cite{mistralai_model}, released by Mistral AI \footnote{Mistral AI - \url{https://mistral.ai/about}} in 2024, a model that stands out for its efficiency and performance despite its compact size, maintaining a balance between result quality and computational efficiency, with an extended context window of 32K tokens; and DistilGPT2 \cite{distilgpt2_model}, a lightweight, optimized version of OpenAI GPT-2 that enables fast and compact responses in resource-constrained environments (such as mobile devices), developed by Hugging Face \footnote{Hugging Face - \url{https://huggingface.co/huggingface}} \cite{sanh2020,xu2024}, offering a balanced performance between speed and response quality. For this selection, we highlight free access to the models, both published under the Apache 2.0 license, and their availability through web platforms and services.
  
    \item \textbf{Main orchestrator}: Coordinates the modules, manages data flow, and ensures the consistency of the certification process.
\end{enumerate}

\subsection{Contributions to Quantum Software Engineering}
The architecture proposed in QAccCert is grounded in and consolidates QSE principles:
\begin{itemize}
    \item \textbf{Separation of concerns}: each module develops a clearly defined functionality through the exposure of a well-specified interface.
    \item \textbf{Hardware abstraction}: the orchestrator does not need to know the details of the FPGA or the quantum backend; it communicates through APIs; this independence between components enables a hardware-agnostic solution through technological decoupling, substantially increasing its deployability across different technologies.
    \item \textbf{Reusability}: modules can be replaced or updated independently (e.g., switching LLM or FPGA).
    \item \textbf{Adaptability}: the system learns and improves with experience, a flexibility requirement that is key to operating in dynamic NISQ environments.
    \item \textbf{Observability}: concrete execution results in conjunction with the generated functional evolution diagrams allow for rapid parametric adjustments to reach optimal configurations.
    \item \textbf{Maintainability}: the modular structure facilitates debugging, testing, and evolutionary software correction.
\end{itemize}

\subsection{Certification Algorithm from the QSE Perspective}
Algorithm~\ref{alg:qacccert} shows the pseudocode of the certification process, highlighting how the components are interconnected following QSE design patterns. The FPGA-accelerated auxiliary CHSH measurement function is shown in Algorithm~\ref{alg:measurechsh}.

\begin{algorithm}[htbp]
\caption{Certification with FPGA and AI}
\begin{algorithmic}[1]
\Require Initial configuration
\Ensure CHSH violation $S$, optimal angles found
\State InitializeHardware() \Comment{FPGA + quantum backend via standard API}
\State $\theta_{\text{current}} \leftarrow [0, \pi/4, \pi/8, 3\pi/8]$ \Comment{Theoretical initial angles}
\State $S_{\text{best}} \leftarrow -\infty$
\State $history \leftarrow []$
\For{\(iter \leftarrow 1\) to \(N_{\text{max}}\)}
    \State \((S, E) \leftarrow \text{MeasureCHSH}(\theta_{\text{current}})\) \Comment{FPGA accelerates E computation}
    \State $history.\text{add}(\{\theta_{\text{current}}, S, E\})$
    \If{$S > S_{\text{best}}$}
        \State $S_{\text{best}} \leftarrow S$
        \State $\theta_{\text{best}} \leftarrow \theta_{\text{current}}$
        \State $\theta_{\text{current}} \leftarrow \text{LocalSearch}(\theta_{\text{current}})$
    \Else
        \If{$iter \mod 2 = 0$}
            \State $\theta_{\text{current}} \leftarrow \text{QueryLLM}(history)$ \Comment{AI optimization}
        \Else
            \State $\theta_{\text{current}} \leftarrow \text{GlobalSearch}(\theta_{\text{best}})$
        \EndIf
    \EndIf
    \If{$|S - 2.828| < \epsilon$}
        \State \textbf{break} \Comment{Practical optimum reached}
    \EndIf
\EndFor
\State \Return $S_{\text{best}}, \theta_{\text{best}}$
\end{algorithmic}
\label{alg:qacccert}
\end{algorithm}

\begin{algorithm}[htbp]
\caption{MeasureCHSH function with FPGA acceleration}
\begin{algorithmic}[1]
\Function{MeasureCHSH}{theta}
    \State Prepare state \(\Psi^- = (|01\rangle - |10\rangle)/\sqrt{2}\)
    \For{{\((\alpha, \beta) \in \{(a,b), (a,b'), (a',b), (a',b')\}\)}}
        \State Measure qubits with angles \(\alpha, \beta\)
        \State Send raw results to FPGA
        \State Compute on FPGA: \(E(\alpha, \beta) = P_{00} + P_{11} - P_{01} - P_{10}\)
    \EndFor
    \State \(S \leftarrow |E(a,b) - E(a,b') + E(a',b) + E(a',b')|\)
    \State \Return \(S, [E(a,b), E(a,b'), E(a',b), E(a',b')]\)
\EndFunction
\end{algorithmic}
\label{alg:measurechsh}
\end{algorithm}

\begin{algorithm}[htbp]
\caption{LLM-based optimization}
\begin{algorithmic}[1]
\Function{QueryLLM}{history}
    \State $Model_{\text{target}}$ $\leftarrow$ LoadModel()
    \State $Iter_{\text{count}}$ $\leftarrow$ GetIterationCount()
    \State $Prompt$ $\leftarrow$ BuildPrompt(history, $Iter_{\text{count}}$) \Comment{Prompt construction from history}
    \State $Response_{\text{JSON}}$ $\leftarrow$ Execute($Model_{\text{target}}$, $Prompt$) \Comment{LLM request and response retrieval}
    \State $\theta_{\text{suggested}}$ $\leftarrow$ ParseAngles($Response_{\text{JSON}}$)
    \State \Return $\theta_{\text{suggested}}$
\EndFunction
\end{algorithmic}
\label{alg:llm}
\end{algorithm}

\subsection{AI-based Optimization: LLM Integration}
The optimization module constitutes the adaptive component of the QSE architecture, responsible for suggesting new measurement angles based on the history of prior iterations. Unlike random search methods that explore the space without memory, the proposed approach uses a Large Language Model (LLM) whose goal is to direct the search toward higher-quality regions of the parameter space, leveraging the history of previous iterations to achieve solutions superior to those obtained through random exploration. See Algorithm~\ref{alg:llm}.

\subsubsection{Optimization Module Architecture}
The optimization module is implemented as a decoupled component that:
\begin{itemize}
    \item Receives from the orchestrator the history of the last $k$ iterations (explored angles and obtained $S$ value).
    \item Constructs a structured prompt encoding this history in natural language.
    \item Queries an LLM (Mistral-7B-Instruct-v0.2 or DistilGPT2) to obtain suggestions for new angles.
    \item Parses the model's JSON response and validates that the angles fall within the range $[0,\pi]$ before sending them to the orchestrator.
\end{itemize}

\subsubsection{Justification for Using LLMs in Our Approach}
The choice of LLMs as the optimization engine is primarily motivated by three inherent characteristics of the problem that make them particularly well-suited:
\begin{enumerate}
    \item \textbf{Exploratory strategy} (intelligent vs. random): pure random search does not leverage information from previous iterations. An LLM, upon receiving the full history, can identify patterns and promising directions, acting as an optimizer with memory.
  
    \item \textbf{Adaptation to imperfect hardware}: on real NISQ devices, the theoretically optimal angles ($\theta = [0,\pi/4,\pi/8,3\pi/8]$) do not coincide with the experimental ones due to noise, temperature, and calibration errors. An LLM can learn from previous iterations on the specific device and suggest specific compensations, whereas random search would treat each iteration as independent, preventing the reuse of useful information and its benefits for future search iterations.
  
    \item \textbf{Correlation between runs}: although each device may exhibit unique and occasional imperfections, common patterns also exist (e.g., certain types of noise). Consequently, an LLM that retains and appropriately uses these data clusters inherent to the sequencing of runs can reuse them as \textit{`search shortcuts'} and accelerate convergence, compared to a more naive strategy such as restarting random searches.
\end{enumerate}

\subsubsection{Model Selection}
We evaluated two models with different performance profiles:
\begin{itemize}
    \item \textbf{Mistral-7B-Instruct-v0.2:} Greater reasoning capability, capable of generating more precise suggestions, but requires a GPU and higher latency ($\sim 500$ ms on optimized hardware).
    \item \textbf{DistilGPT2:} A lighter model with low resource demands, executable on a CPU, with reduced latency ($\sim 100$ ms), but potentially generating less sophisticated suggestions. \\
    Both models were tested under standard parameterizations with temperature 0.5.
\end{itemize}

Both models demonstrate the ability to suggest promising angles, achieving a more intelligent exploration of the parameter space that translates into higher-quality solutions compared to random search, though without a significant reduction in the number of iterations in this experiment. Mistral can reach slightly higher $S$ values in preliminary tests, albeit with greater computational cost. We highlight that our modular architecture promotes an experimental system with reduced coupling and well-cohesive independent components, allowing model swapping.

\section{Experimental Results}
All results presented in this section were obtained through \textbf{quantum simulations} using Qiskit Aer (AerSimulator \footnote{Qiskit - AerSimulator \url{https://qiskit.github.io/qiskit-aer/stubs/qiskit_aer.AerSimulator.html}}) with a finite number of shots (shot-based simulation), without an incorporated noise model. The purpose of these simulated experiments is to validate the proposed hybrid architecture (QAccCert), quantify the potential benefit of FPGA acceleration and LLM-guided optimization, and demonstrate the conceptual feasibility of the approach before its application on real NISQ hardware.
To validate the hypothesis that FPGA+AI interconnection, guided by QSE principles, is necessary for practical certification, we evaluated the impact of two essential factors: (i) hardware acceleration via FPGA in the computation of correlations, and (ii) AI-guided algorithmic optimization. To this end, we compare three optimization strategies that benefit equally from the base acceleration provided by the FPGA:
\begin{enumerate}
    \item \textbf{Pure random search (baseline)}: Generates random measurement angles at each iteration, without using historical information. This strategy establishes the baseline for algorithmic efficiency.
  
    \item \textbf{Fallback (local perturbation)}: In the absence of the LLM, the system uses a fallback mechanism that explores the neighborhood of the best angle found so far through random perturbations. This strategy allows isolating the specific benefit of AI.
  
    \item \textbf{LLM optimization}: The AI module analyzes the history of previous iterations and suggests new measurement angles, guiding the search toward promising regions of the parameter space.
\end{enumerate}
It is worth noting that all three strategies execute the same correlation computation core. This makes it possible to separately analyze:
\begin{itemize}
    \item \textbf{Algorithmic advantage}: Improvement in solution quality (S value) and reduction in iterations thanks to AI (Table~\ref{tab:convergencia}).
    \item \textbf{Hardware advantage}: Acceleration of time per iteration and scalability via FPGA (Tables~\ref{tab:resultados_fpga} and~\ref{tab:aceleracion_hardware}).
\end{itemize}

\subsection{QSE Evaluation Metrics}
To quantify the benefits of the proposed architecture, we define the following metrics based on Quantum Software Engineering principles:
\begin{itemize}
\item \textbf{Certification quality}: The CHSH violation value $S$ achieved and its proximity to the theoretical limit $2\sqrt{2} \approx 2.828$. This effectiveness metric evaluates the precision of the solution found (Table~\ref{tab:convergencia}).
  
\item \textbf{Computational efficiency}: Time per iteration and total time required to reach convergence. This performance metric allows comparing the computational cost of each strategy (Table~\ref{tab:aceleracion_hardware}).
  
\item \textbf{Hardware acceleration}: Comparison of the pure computation time for correlations on PC vs. FPGA, expressed as speedup (time ratio). This metric quantifies the intrinsic gain of hardware acceleration (Table~\ref{tab:resultados_fpga}).
  
\item \textbf{Throughput}: Number of correlation computations per second on each platform, a key metric for evaluating scalability in high-demand scenarios.
  
\item \textbf{Adaptability}: The AI module's ability to guide the search toward higher-quality regions of the parameter space, measured by the improvement in the achieved $S$ value compared to random search. This learning metric reflects the qualitative advantage of the LLM: it does not necessarily reduce the number of iterations, but rather makes better use of them to reach superior solutions (Table~\ref{tab:convergencia}).
  
\item \textbf{Robustness}: Standard deviation of $S$ across runs, indicating the reproducibility and reliability of the certification process.
\end{itemize}

\subsection{Comparative Results}
To evaluate the impact of hardware acceleration and AI-based optimization, we conducted a systematic comparison along two dimensions: (i) FPGA performance in the computation of correlations, and (ii) the efficiency of different optimization strategies that benefit from this acceleration.

\subsubsection{FPGA Performance in Correlation Computation}
Table~\ref{tab:resultados_fpga} presents a multi-level performance analysis of the FPGA, decomposing times to identify bottlenecks and future potential. This analysis enables understanding of the real acceleration capability of the hardware, independently of the optimization strategy employed.

\begin{table}[htbp]
\caption{Multi-level FPGA performance analysis}
\label{tab:resultados_fpga}
\vspace{-0.5cm}
\begin{center}
\footnotesize
\renewcommand{\arraystretch}{1.2}
\setlength{\tabcolsep}{3pt}
\begin{tabular}{|c|c|c|}
\hline
\textbf{Level} & \textbf{Metric} & \textbf{Value} \\
\hline
\multirow{2}{*}{PC Only} & PC Time & $0.00197 \pm 0.0003$ ms \\
 & per S computation & \\
\hline
\multirow{2}{*}{\makecell{FPGA+UART \\ \textit{(Current prototype)}}} & Time/iteration & $554.4 \pm 5.2$ ms \\
 & Breakdown & \makecell{$<0.1\%$ computation \\ + $99.9\%$ delay} \\
\hline
\multirow{2}{*}{\makecell{FPGA+USB 3.0 \\ \textit{(Projected)}}} & Time/iteration & $0.00084$ ms \\
 & Scalable throughput & depending on parallelization \\
 \hline
\end{tabular}
\end{center}
\footnotesize
\textbf{Note:} The PC Only level shows pure computation time: PC 0.00197 ms per correlation, FPGA 0.00083 ms for 4 correlations in parallel. The prototype uses UART with intentional delays and no parallelization (550 ms). The projected level assumes a USB 3.0 interface (5 Gbps) with a communication time of 0.00001 ms, resulting in 0.00084 ms per complete iteration (0.00083 ms computation + 0.00001 ms communication).
\end{table}

The reported iteration times for the FPGA (average 554.4 ms) correspond to the current prototype and include software delays deliberately introduced to ensure communication integrity: 20 ms initial delay, 4 pauses of 10 ms between byte transmissions, and 500 ms response wait. This distinction is paramount: the observed overhead is a characteristic of the test software and the UART interface, not a limitation of the hardware acceleration, whose projected speedup with USB 3.0 reaches 9.4x over the CPU and enables a throughput scalable to thousands of correlations in parallel.

\subsubsection{Optimization Strategy Efficiency}
Building on this hardware acceleration baseline, we evaluated the three optimization approaches that determine solution quality and the number of iterations required. Table~\ref{tab:convergencia} compares these strategies in terms of precision (achieved S value) and algorithmic efficiency (iterations to the maximum).

\begin{table}[htbp]
\caption{Convergence efficiency: comparison of optimization strategies for CHSH certification}
\label{tab:convergencia}
\begin{center}
\footnotesize
\renewcommand{\arraystretch}{1.2}
\setlength{\tabcolsep}{2pt}
\begin{tabular}{|c|c|c|c|}
\hline
\textbf{Optimization} & \textbf{\(S\)} & \textbf{Iteration} & \textbf{Efficiency\textsuperscript{a}} \\
\textbf{strategy} & & \textbf{of maximum} & \\
\hline
Pure random search & $2.7520$ & $42$ & $97.3\%$ \\
\hline
Fallback (local perturbation) & $2.7686$ & $35$ & $97.9\%$ \\
\hline
LLM optimization & & & \\
\quad Distil GPT-2 & $2.8267$ & $41$ & $99.94\%$ \\
\quad Mistral-7B & $2.6909$ & $46$ & $95.1\%$\\
\hline
\end{tabular}
\end{center}
\footnotesize
\textsuperscript{a} Efficiency = $(S$ achieved $/$ $2\sqrt{2} \approx 2.828$) $\times 100$\%. \\
\textbf{Note:} All $S$ values, number of iterations, and convergence were obtained through quantum simulations using Qiskit's \textbf{AerSimulator} (without a noise model). On real NISQ devices, CHSH violation values will be lower.
\end{table}

Table~\ref{tab:aceleracion_hardware} shows the real-time impact of running these same strategies on FPGA hardware with the projected performance.

\begin{table}[htbp]
\caption{Hardware acceleration: comparison of total execution times}
\label{tab:aceleracion_hardware}
\begin{center}
\footnotesize
\renewcommand{\arraystretch}{1.2}
\setlength{\tabcolsep}{2pt}

\begin{tabular}{|c|c|c|c|c|}
\hline
\textbf{Optimization} & \textbf{Iter} & \textbf{Total time} & \textbf{Total time} & \textbf{Total time} \\
 \textbf{strategy} & & \textbf{CPU} & \textbf{FPGA+UART} & \textbf{FPGA+USB 3.0} \\
 & & \textbf{(serial)} & \textbf{(prototype)} & \textbf{(parallel computation)} \\
\hline
Random search & 42 & 0.331 ms & 23,285 ms & 0.0353 ms \\
\hline
Fallback & 35 & 0.276 ms & 19,404 ms & 0.0294 ms \\
\hline
LLM optimization & & & & \\
Distil GPT-2 & 41 & 0.323 ms & 22,730 ms & 0.0344 ms \\
Mistral-7B & 46 & 0.362 ms & 25,502 ms & 0.0386 ms \\
\hline
\end{tabular}
\end{center}
\footnotesize
\textbf{Note:} The purely hardware speedup is 9.4× (time per iteration CPU vs FPGA+USB 3.0) and is constant regardless of strategy.
\end{table}

The results obtained, summarized in Tables~\ref{tab:convergencia} and~\ref{tab:aceleracion_hardware}, validate the central thesis of this work. The analysis is broken down into three fundamental aspects: precision, algorithmic efficiency, and hardware acceleration.
\begin{itemize}
    \item \textbf{Precision (Solution Quality):} Certification quality improves significantly with the optimization strategy. Pure random search achieves a value of $S = 2.7520$ (97.3\% of the theoretical limit). The fallback (local perturbation) improves slightly to $S = 2.7686$ (97.9\%). However, LLM optimization achieves a value of $\mathbf{S = 2.8267}$, representing an \textbf{efficiency of 99.94\%} with respect to the theoretical maximum of $2.828$, demonstrating the LLM's capability to guide the search toward optimal regions of the parameter space.
    \item \textbf{Algorithmic Efficiency (Convergence Speed):} Regarding the number of iterations required to reach the maximum, the fallback is the fastest, converging in only 35 iterations (17\% fewer than random). The LLM converges in 41 iterations, while random search requires 42. This reveals a \textbf{trade-off}: the LLM's advantage is qualitative (solution quality), not quantitative (number of iterations); the fallback prioritizes speed, quickly finding a local maximum, while the LLM invests a few additional iterations to perform a more intelligent exploration that allows it to reach a superior solution.
   
    \item \textbf{Hardware Acceleration with FPGA:} The hardware advantage becomes apparent when implementing the optimization strategies on the FPGA. Using a USB 3.0 interface (5 Gbps), the time per iteration on the FPGA is reduced to 0.00084 ms (parallel computation 0.00083 ms + communication 0.00001 ms). As shown in Table~\ref{tab:aceleracion_hardware}, for LLM optimization (41 iterations), the FPGA completes processing in 0.0344 ms, while the CPU requires 0.323 ms for the same task.

\begin{figure}[htbp]
\centerline{\includegraphics[width=0.9\columnwidth]{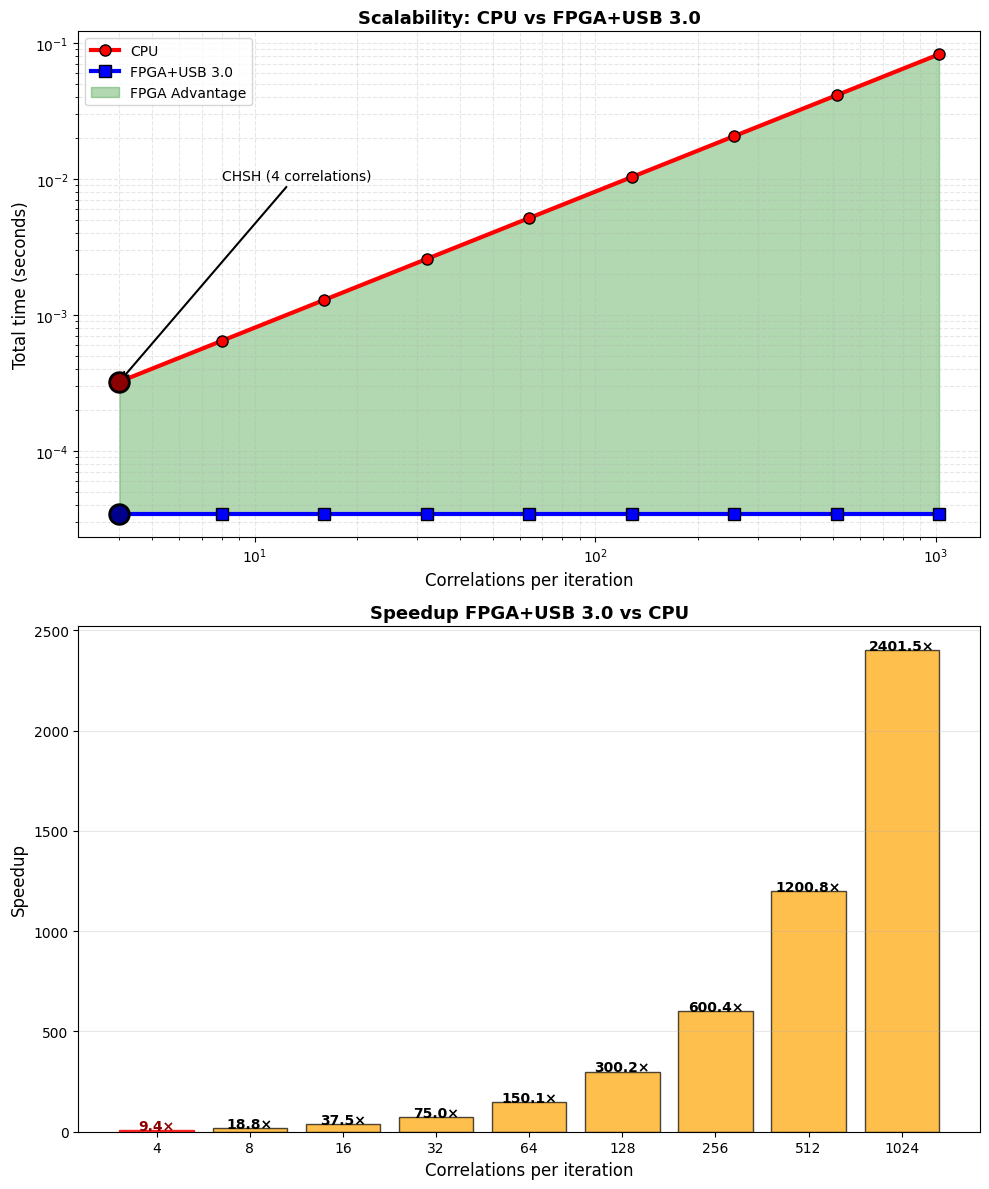}}
\caption{FPGA acceleration scalability. Top: Total execution time for 41 iterations on a logarithmic scale. The CPU (red) scales linearly with the number of correlations ($t_{CPU} = n_{corr} \times 41 \times 0.00197$ ms), while FPGA+USB 3.0 (blue) maintains a constant time ($t_{FPGA} = 41 \times 0.00084$ ms) thanks to parallel processing. The green area represents the cumulative FPGA advantage. Bottom: Resulting speedup, from \textbf{9.4×} for CHSH (4 correlations) to \textbf{$>$2,400×} for systems with 1024 correlations.}
\label{fig:aceleracion_fpga}
\end{figure}
\end{itemize}

\begin{figure}[htbp]
\centerline{\includegraphics[width=0.7\columnwidth]{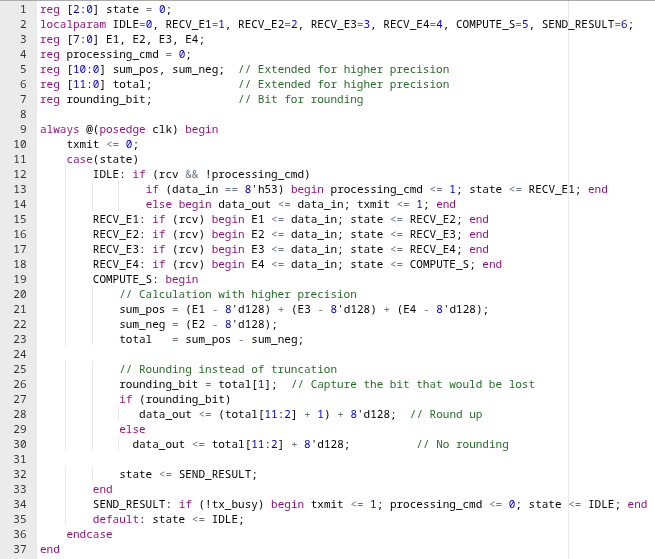}}
\caption{Verilog code snippet for correlation computation on the FPGA.}
\label{fig:verilog}
\end{figure}

Figure~\ref{fig:verilog} shows a snippet of the \textbf{Verilog} \footnote{Verilog Hardware Description Language - \url{https://www.verilog.com/}} code implemented on the FPGA, evidencing hardware-level optimization as part of the QSE architecture.

\section{Discussion}
\subsection{Lessons for Quantum Software Engineering}
The results lead us to believe that the technology combination guided by QSE is not a luxury, but a practical necessity. Specifically:
\begin{itemize}
    \item Modularity enables independent evolution: We could replace the LLM with another model or the FPGA with a GPU without redesigning the entire system.
    \item Hardware abstraction facilitates testing: We can test each module separately with simulators before real deployment.
    \item Adaptability improves with experience: The system learns from previous runs and continuously optimizes itself.
    \item Quantifiable metrics enable objective evaluation: The violation $S$, convergence time, and standard deviation provide a basis for comparing design alternatives.
\end{itemize}
In simulations with few iterations, LLM latency may be negligible. However, in real scenarios with noisy quantum hardware, where each iteration is costly (both in time and resources), the reduction in the number of iterations may justify the use of AI. A typical design trade-off example in software engineering.

\subsection{Observed Limitations on Real NISQ Hardware}
On real NISQ hardware, experimental CHSH violations typically range between 2.3 and 2.7 (depending on the device, number of shots, error mitigation, and connectivity), even with error mitigation techniques or zero-noise extrapolation \cite{waring2025}, well below the theoretical limit. The results presented here represent an ideal upper bound and serve as a reference for comparing future implementations with mitigation.
Although the experiments were conducted in simulations (without a noise model), it is important to contextualize the results with the reality of current NISQ devices. In superconducting or trapped-ion processors, accumulated noise drastically limits circuit depth and entanglement quality. Therefore, the QAccCert framework, with its focus on adaptive optimization via LLM and FPGA acceleration, could be especially valuable in real scenarios to compensate for these imperfections through more efficient searches and fast data processing.

The current evidence suggests that quantum computing will not be used in isolation, but integrated with classical computers \cite{pousa2024}, as it will be necessary to prepare the execution of quantum algorithms beforehand in classical computing and subsequently process the results classically. This post-processing could require enormous computational power and, given current hardware limitations, it would be infeasible to perform it in a reasonable time using a single computer sequentially. The natural and widely adopted response is the use of HPC (High Performance Computing), which is based on parallel programming to accelerate computation. For this reason, the integration of QC systems with HPC infrastructures is proposed \cite{pousaQPUs2024}, typically combining multicore processors, GPUs, or FPGAs (as in our case) in the same machine and even multiple interconnected machines forming clusters that pool their computational power. Additionally, artificial intelligence techniques can be employed to facilitate parameter optimization in hybrid quantum-classical algorithms, reducing the complexity of their implementation.

\section{Future Work and Conclusions}
\subsection{Implications for Quantum Software Engineering}

This work demonstrates that the practical certification of quantum computers requires a software engineering approach that integrates:
\begin{itemize}
    \item Modular architectures with clear separation of concerns.
    \item Hardware abstraction through well-defined APIs.
    \item Adaptive components (AI) that learn from experience.
    \item Hardware acceleration by parallelizing computation.
    \item Quantifiable metrics for objective evaluation.
\end{itemize}
QAccCert serves as a case study validating these principles, showing concrete computational results using entanglement certification through CHSH \cite{Arango}. This work joins other recent efforts \cite{lammers2025qonscious, qctrl_fireopal} that, from different perspectives, contribute to the goals of Quantum Software Engineering by:
\begin{itemize}
    \item Providing concrete examples of the application of QSE principles.
    \item Demonstrating the feasibility of hybrid architectures in real scenarios.
    \item Identifying metrics and methodologies for evaluating quantum systems.
    \item Opening research questions on the design, maintenance, and evolution of such systems.
\end{itemize}

\subsection{Optimization with a Theoretical Starting Point}
As a future line of work, we plan to explore hybrid strategies that combine:
\begin{itemize}
    \item Initialization with theoretical angles: Beginning the search from the ideal optimal angles ($\theta = [0,\pi/4,\pi/8,3\pi/8]$) rather than from random starting points.
    \item LLM refinement: Using the LLM to adjust these angles to the specific hardware imperfections, further reducing the number of iterations needed.
    \item LLM fine-tuning: Training the model with historical data from multiple runs on the same device to improve the quality of suggestions.
\end{itemize}
This combination is expected to reduce the number of iterations to fewer than 40, approaching the efficiency of a specialized optimizer but with the flexibility of an AI-based system.

\subsection{Future Research Directions in QSE}
\begin{itemize}
    \item \textbf{Scalability to multipartite systems}: Extending the approach to certification of GHZ states and other complex quantum resources, evaluating how the QSE architecture adapts to new protocols.
    \item \textbf{Hardware-software optimization}: Implementing more functionality directly on the FPGA (e.g., random number generation, experiment control) and studying the optimal partitioning between hardware and software, drawing inspiration from efficient simulation techniques such as those based on graph states \cite{PhysRevA.73.022334}.
    \item \textbf{AI improvements for QSE}: Incorporating reinforcement learning for continuous optimization, fine-tuning LLMs with specific experimental data, or other techniques for greater precision in angle suggestions, complementing invariant estimation methods \cite{Holik2024GroupinvariantEO}.
    \item \textbf{Integration with other frameworks}: Exploring the complementarity between QAccCert and proposals such as Qonscious \cite{lammers2025qonscious} or commercial solutions like Fire Opal \cite{qctrl_fireopal}, as well as resource estimation tools like BenchQ \cite{benchq_github} and Rigetti Resource Estimation \cite{rigetti_github}, to create more reliable quantum software ecosystems.
    \item \textbf{Requirements engineering for quantum systems}: Developing methodologies to specify, validate, and verify requirements for hybrid quantum-classical software.
    \item \textbf{Quality and testing in QSE}: Defining quality metrics specific to quantum software and developing automated testing tools.
    \item \textbf{Containerization and continuous deployment}: Facilitating framework adoption through containers, continuous integration, and reproducible environments.
\end{itemize}
The current prototype uses UART communication at 115200 baud, with intentional synchronization delays that ensure robust transmission but increase the time per iteration (~550 ms). It is important to note that this overhead is a characteristic of the instrumentation software, not a limitation of the hardware acceleration: the pure computation time on the FPGA could be less than 1 µs per correlation.
In future versions, the UART interface could be replaced with higher-speed alternatives, such as USB 3.0 (up to 5 Gbps) or PCIe (several GB/s), reducing communication overhead and allowing FPGA acceleration to be fully leveraged. For HPC environments or quantum data centers, the integration of FPGAs with high-speed optical interconnects is envisioned, which would enable remote control of quantum experiments with minimal latency and high scalability. This evolution would enable distributed architectures.

\section{Conclusion}
The NISQ era demands certification methods that go beyond ideal assumptions. This paper has argued and demonstrated that the principles of classical software engineering are applicable and valuable in the quantum domain as well. Experimentally, we have confirmed that QSE strategies combined with the interoperability of FPGAs and AI models are useful for addressing this challenge in a systematic manner.

Specifically, the results obtained validate the three dimensions of the proposed approach: LLM optimization achieves a CHSH violation of $S=2.8267$ (99.94\% of the theoretical limit), demonstrating its capacity for efficient parameter space exploration; the local perturbation fallback mechanism reduces the required iterations by up to 17\%, optimizing convergence speed; and FPGA acceleration projects superior speedups in high-dimensionality scenarios due to parallelization capability. These contributions are complementary and confirm the viability of the hybrid approach, whose integration with HPC infrastructure also enables meeting the scalability requirements inherent to real quantum environments.

The main contribution is not the framework itself, but the demonstration that the synergy between hardware acceleration and artificial intelligence, framed within a well-designed software architecture, transforms quantum certification from a theoretical problem into a practical reality. This principle, which is beginning to be explored by multiple initiatives both academic \cite{lammers2025qonscious} and commercial \cite{qctrl_fireopal}, is extensible to other areas of quantum computing where adaptation to imperfect hardware and computational efficiency are critical.

\bibliographystyle{IEEEtran}
\bibliography{referencias}
\end{document}